\documentclass[epjST]{svjour}
\usepackage{graphics}
\usepackage{graphicx}
\usepackage{amsmath}
\usepackage{amssymb}
\usepackage{hyperref}
\usepackage{subfigure}
\begin{document}
\title{Generalized Synchronization of Coupled Chaotic Systems}
\author{Suman Acharyya\thanks{\email{suman@prl.res.in}} \and R. E. Amritkar\thanks{\email{amritkar@prl.res.in}}}
\institute{Physical Research Laboratory, Ahmedabad, India.}

\abstract{
In this paper we briefly report some recent developments on generalized synchronization. We discuss different methods of detecting generalized synchronization. We first consider two unidirectionally coupled systems and then two mutually coupled systems. We then extend the study to a network of coupled systems. In the study of generalized synchronization of coupled nonidentical systems we discuss the Master Stability Function (MSF) formalism for coupled nearly identical systems. Later we use this MSF to construct synchronized optimized networks. In the optimized networks the nodes which have parameter value at one extreme are chosen as hubs and the pair of nodes with larger difference in parameter are chosen to create links.
} 
\maketitle
\section{Introduction}
\label{intro}

Study of synchronization of coupled chaotic system has attracted much attention recently~\cite{Strogatz2008,Pikovsky2001,Boccaletti2002,Arenas2008}. Synchronization of coupled dynamical systems can be defined as a process where two or more coupled systems adjust their trajectories to a common behavior. In the literature, the most commonly studied synchronization is between systems which have exactly the same or identical dynamical equations. Two identical coupled chaotic systems are said to be synchronized when the state variables of the coupled systems become equal. This type of synchronization is known as \emph{complete synchronization}(CS)~\cite{Fujisaka1983,Afraimovich1986,Pecora1990}. However for coupled nonidentical systems it not possible to observe CS, instead one will find other form of synchronization such as \emph{phase synchronization}(PS)~\cite{Rosenblum1996}, \emph{generalized synchronization}(GS)~\cite{Abarbanel1995}, etc. PS is a weaker form of synchronization, where the phases of the coupled 
systems become locked but their amplitudes are in general unrelated. In GS the state variables of the coupled systems are related by some function. GS occurs mainly for coupled nonidentical systems. The CS can be consider as special cases of GS. In this paper we will consider GS of coupled chaotic systems.

\section{Generalized synchronization for two coupled chaotic systems}
\label{gen-sync-2-uni-osc}

\begin{figure}
\centering
\includegraphics[width=.9\columnwidth]{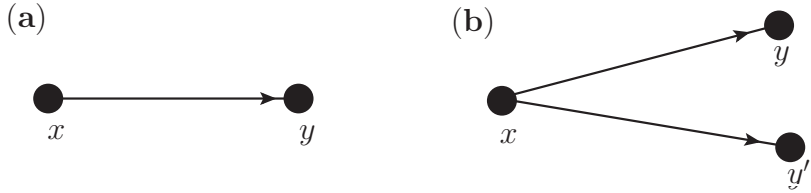}
\caption{(a)Schematic diagram of a drive-response configuration. Here, the system $x$ is driving the system $y$.(b) The schematic diagram of auxiliary system approach is shown here. $y'$ is a replica of the response system $y$ and is driven by the same system $x$.}
\label{schematic-drive-response}
\end{figure}

In this section we review generalized synchronization between two coupled chaotic systems.
The concept of GS was introduced first for unidirectionally coupled systems, i.e. systems coupled in a drive response configuration, by Abarbanel et. al.~\cite{Abarbanel1995} (see Fig.~\ref{schematic-drive-response}(a)). Let us consider a $d$ dimensional system $x$ driving a $r$ dimensional system $y$. The dynamics of the drive and the response systems can be written as
\begin{subequations}
\label{dynamical-equations}
\begin{eqnarray}
\dot{x}(t) &=& F(x(t)) \label{dyn-eqn-x}\\
\dot{y}(t) &=& G(y(t)) + \varepsilon H(x(t),y(t))\label{dyn-eqn-y}
\end{eqnarray}
\end{subequations}
where $x(\in R^d)$ and $y(\in R^r)$ are the state variables of drive and response systems respectively, $\varepsilon$ is the coupling parameter, $F: R^d \rightarrow R^d$ and $G: R^r \rightarrow R^r$ give the uncoupled dynamics of the drive and response systems respectively, and $H: R^d \oplus R^r \rightarrow R^r$ is the driving function. For a suitable driving function $H$ and sufficiently large coupling parameter $\varepsilon$, systems (\ref{dyn-eqn-x}) and (\ref{dyn-eqn-y}) can exhibit GS. 

When $\varepsilon=0$ the evolution of the response system is independent of the drive system. As the coupling parameter $\varepsilon$ is increased the coupled systems are said to show generalized synchronization when there exists a map $\phi: R^d \rightarrow R^r$ relating the state variables of the response to that of the drive systems, i.e.
\begin{equation} 
y(t)=\phi(x(t)).
\label{func-relation}
\end{equation}
The synchronization manifold is defined by the condition $y(t) = \phi(x(t))$ and the motion of the synchronized systems will collapse onto this synchronization manifold. For most cases it is difficult to determine the functional relation between the coupled systems. For GS this functional relation must be observed for the trajectories on the attractors, but not necessarily for the transient trajectories.

\section{Detection of Generalized Synchronization}
\label{detect-gen-sync}

In this section we will discuss the schemes that have been developed to detect generalized synchronization.

\subsection{Mutual False Nearest Neighbors (MFNN) method}
\label{MFNN-method}
In this section we briefly discuss the mutual false nearest neighbors (MFNN) method for detecting generalized synchronization~\cite{Abarbanel1995}. The main feature of this method is the concept of local neighborliness. In the generalized synchronized state the trajectories of the drive and response systems are connected by a functional relationship~(\ref{func-relation}).
This method depends on the observation that in the synchronized state, two neighboring  points in the phase space of drive system correspond to two neighboring points in the phase space of response system. 

Let us consider a time series of the drive variable $x_1, x_2, \ldots, x_T$ and the corresponding time series of the response variable $y_1, y_2, \ldots, y_T$. From the time series the attractors of the drive and the response systems can be reconstructed using embedding methods~\cite{Farmer1980,Abarbanel1992}. Let the dimension of the drive and the response systems are $d_d$ and $d_r$ respectively and each are larger than the respective global embedding dimensions required to unfold the attractors. Choose  an arbitrary point $x_n$ from the drive time series. Let the nearest phase space neighbor of this point in the time series be $x_{n_{NND}}$. In the generalized synchronization, we can expect that the corresponding points of the response system $y_n$ and $y_{n_{NND}}$ are close. Using Eq.~(\ref{func-relation}), the distance between these two points of the response system can be written as
\begin{equation}
y_n - y_{n_{NND}} = \phi(x_n) - \phi(x_{n_{NND}}). \label{response-distance}
\end{equation}
As the difference is expected to be small, we can write Eq.~(\ref{response-distance}) as,
\begin{equation}
y_n - y_{n_{NND}} = D \phi(x_n) (x_n - x_{n_{NND}}) \label{response-distane-linearized-1}
\end{equation}
where $D\phi(x_n)$ is the Jacobian matrix evaluated at $x_n$.

Similarly, now we consider the point $y_n$ from the phase space of the response system and find its nearest neighbor from the time series as $y_{n_{NNR}}$. Let the  corresponding points of the drive system be $x_n$ and $x_{n_{NNR}}$. Using Eq.~(\ref{func-relation}), the distance between the points of the response variable can be written as
\begin{equation}
y_n - y_{n_{NNR}} = D \phi(x_n) (x_n - x_{n_{NNR}}) \label{response-distane-linearized-2}
\end{equation}
The MFNN parameter is defined as the ratio,
\begin{equation}
p = \frac{1}{T} \sum_n \frac{|y_n - y_{n_{NND}}| \ |x_n - x_{n_{NNR}}|}{|x_n - x_{n_{NND}}| \ |y_n - y_{n_{NNR}}|}.
\label{MFNN-param}
\end{equation}
In the synchronized state the MFNN parameter will be order of unity.

\subsection{Auxiliary System Approach (ASA)}
\label{ASA-unidir}

Auxiliary system approach (ASA) is another way of detecting generalized synchronization between the drive system $x$ and the response system $y$ of Eq.~(\ref{dynamical-equations})~\cite{Abarbanel1996}. In ASA, we consider another replica of the response system, $y'$, driven by the same system $x$ (see Fig.~\ref{schematic-drive-response}(b)) and the dynamics of this auxiliary system $y'$ is given as
\begin{equation}
\dot{y}'(t) = G(y'(t)) + \varepsilon(x(t) - y'(t))
\label{aux-sys-y'}
\end{equation}
where $\varepsilon$ is the coupling constant.

When system $y$ is not in generalized synchronization with system $x$, then the trajectories of the response system and the auxiliary system will be unrelated. When $x$ and $y$ are in generalized synchronization with the relation $y(t)=\phi(x(t))$, then there clearly exists a solution $y(t)=y'(t)$ with $y'(t)=\phi(x(t))$. The stability of the synchronization manifold $y(t)=\phi(x(t))$ ensures that $y'(t)$ can track $y(t)$ as $y'(t)=y(t)$.  Thus, in the case of generalized synchronization the orbits of the response system and the auxiliary system tend to each other after the transients die out, i.e. $y'(t) \to y(t)$.

It can be shown that the linear stability of the manifold $y'(t) = y(t)$ is the same as the linear stability of the synchronization manifold $y(t) = \phi(x(t))$. To see this let us consider the linearized equations in $z(t) = y(t) - \phi(x(t))$ and $z'(t) = y'(t) = \phi(x(t))$ as
\begin{eqnarray}
\dot{z}(t) &=& DG(\phi(x(t))).z(t) - \varepsilon z(t) \label{response-lin-eq}\\
\dot{z}'(t) &=& DG(\phi(x(t))).z'(t) - \varepsilon z'(t) \label{auxiliary-lin-eq}
\end{eqnarray}
where $DG$ is the Jacobian matrix calculated at the synchronized solution $\phi(x(t))$. Since, the linearized equations for $z(t)$ and $z'(t)$ are identical, the linearized equations for $z'(t) - z(t) = y'(t) - y(t)$ is also identical to them, i.e.
\begin{equation}
(\dot{z}'(t) - \dot{z}(t)) = DG(\phi(x(t))) (z'(t) - z(t)) - \varepsilon (z'(t) - z(t)).
\label{lin-eq-res-aux}
\end{equation}
Therefore if the manifold of the generalized synchronized motion $y(t) = \phi(x(t))$ is stable then the manifold is linearly stable for $y'(t)-y(t)$ and vice versa.

\begin{figure}
\centering
\includegraphics[width=.9\columnwidth]{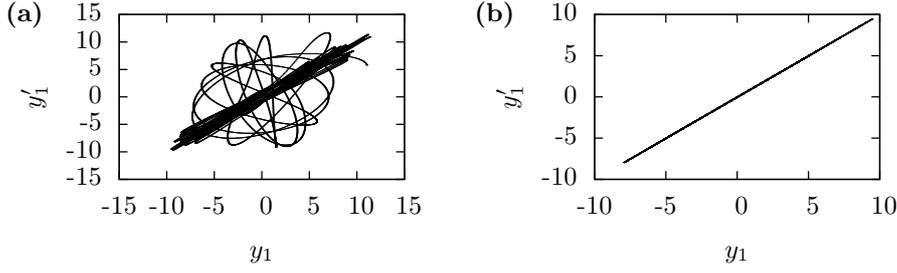}
\caption{The projection of the phase space trajectory of the response R\"ossler system and its auxiliary system is plotted in the $y_1-y_1'$ plane for (a) no synchronization, $\varepsilon=0.15$, and (b) generalized synchronization, $\varepsilon=0.20$. The parameter are $\omega_x=1.015, \, \omega_y=0.985$ and $a_r=b_r=0.2, \, c_r=7.0$.}
\label{ros_phase_space_projection}
\end{figure}

As an example, we consider two nonidentical chaotic R\"ossler systems coupled in drive response configuration. The drive R\"ossler system is
\begin{eqnarray}
\dot{x}_1 &=& -\omega_x x_2 - x_3, \nonumber\\
\dot{x}_2 &=& \omega_x x_1 + a_r x_2, \label{rossler-drive}\\
\dot{x}_3 &=& b_r + x_3(x_1 - c_r), \nonumber
\end{eqnarray}
and the response R\"ossler systems is
\begin{eqnarray}
\dot{y}_1 &=& -\omega_y y_2 - y_3 + \varepsilon(x_1 - y_1),\nonumber\\
\dot{y}_2 &=& \omega_y y_1 + a_r y_2, \label{rossler-response}\\
\dot{y}_3 &=& b_r + y_3(x_1 - c_r), \nonumber
\end{eqnarray}
where $\omega_{x,y}, a_r, b_r, c_r$ are R\"ossler parameters with $\omega_x \neq \omega_y$. For this configuration GS can be observed for large coupling parameters. To test the ASA we make an auxiliary system $y'$ of the response system $y$ and drive this replica in exactly the same way as the response system. In Fig.~\ref{ros_phase_space_projection}, the projection of the post transient phase space trajectory of the response R\"ossler system and the auxiliary system is plotted in the $y_1-y_1'$ plane for (a) no synchronization and (b) GS.

\subsection{Lyapunov Exponents Method}
\label{lyap_expo}

Here we briefly discuss the Lyapunov exponent method to analyze GS of coupled chaotic systems. The LE calculation is known to detect the GS boundary more precisely than MFNN and ASA methods. MFNN and ASA give mainly the qualitative confirmation of GS regime. The criteria for stable CS in unidirectionally coupled identical systems is that all the Lyapunov exponents which are transverse to the synchronization manifold are negative~\cite{Vaidya1992}. These Lyapunov exponents which are transverse to the synchronization manifold are known as the \emph{transverse Lyapunov exponents} (TLE).

One can extend the same idea to analyze stability of GS for coupled nonidentical systems~\cite{Parlitz1996}. In this case, GS occurs if and only if all the transverse Lyapunov exponents are negative.

We consider the drive-response configuration given by Eq.~(\ref{dynamical-equations}). Here the dimension of the drive system is $d$ and the response system is $r$. The behavior of these unidirectionally coupled system is characterized by the Lyapunov exponent spectrum of the entire system. For this configuration the drive systems will evolve independently thus its LE spectrum will not be affected by the response system. Let $\lambda_1\geqslant\lambda_2\geqslant...\geqslant\lambda_{(d+r)}$ give the LE spectrum of the entire coupled system. Of these LEs it is easy to identify those corresponding to the drive system, since in the corresponding eigenvectors the components corresponding to the response system are zero. The remaining exponents are the transverse Lyapunov exponents (TLEs). The GS is stable when all the TLEs are negative or equivalently the largest TLE is negative.

\begin{figure}
\centering
\includegraphics[width=.9\columnwidth]{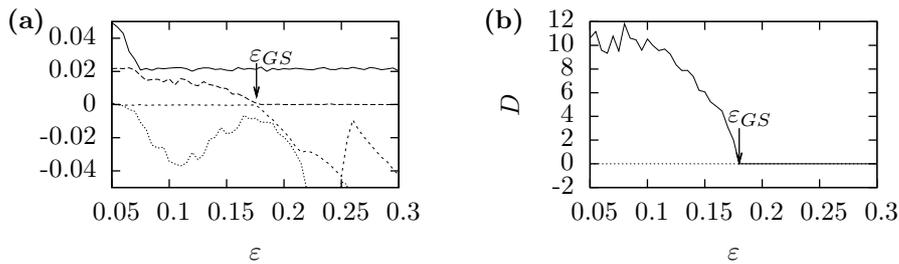}
\caption{(a) The four largest Lyapunov exponents for two unidirectionally coupled R\"ossler systems are shown as a function of the coupling parameter $\varepsilon$. As the coupling parameter is increased the coupled system first undergo phase synchronization when one zero exponent become negative. At couping parameter $\varepsilon=\varepsilon_{GS}\sim0.175$, one positive exponent become negative and the coupled systems undergo generalized synchronization. (b) The time averaged Euclidean distance between the response system and the auxiliary system is plotted as a function of coupling parameter $\varepsilon$. At $\varepsilon=\varepsilon_{GS}$, the distance tends to zero and the coupled systems undergo generalized synchronization. The parameters of the coupled R\"ossler systems are the same as in Fig.~\ref{ros_phase_space_projection}.}
\label{err_res_aux_ale_ros_uni}
\end{figure}

Let us consider two unidirectionally coupled nonidentical R\"ossler systems as given in Eq.~(\ref{rossler-drive}) and Eq.~(\ref{rossler-response}). In Fig.~\ref{err_res_aux_ale_ros_uni}(a), the four largest Lyapunov exponents of the coupled systems are shown as a function of coupling parameter $\varepsilon$. For $\varepsilon=0.0$, the figure shows two positive and two zero exponents. As coupling parameter $\varepsilon$ is increased first one zero exponent become negative at this point phase synchronization occurs between the coupled systems~\cite{Rosenblum1996}, with further increase in coupling parameter one positive exponent become negative at $\varepsilon=\varepsilon_{GS}\sim 0.175$ and at this point the coupled systems undergo generalized synchronization~\cite{Parlitz1996}. For comparison with ASA we plot the time average Euclidean distance, $D=1/T\sum_t \parallel y(t) - y'(t) \parallel$, between the response system and its auxiliary system as a function of the coupling parameter $\varepsilon$ in Fig.~\ref{err_res_aux_ale_ros_uni}(b) which shows that the distance between the response and the auxiliary system tends to zero at $\varepsilon_{GS}\sim 0.175$.

\section{Generalized synchronization for mutually coupled systems}
\label{gen-sync-mutual}

In the above section we have discussed the emergence and detection of GS for unidirectionally coupled systems. We have seen that in GS for unidirectionally coupled systems  there exists a functional relation between the state variables of the drive and response systems, i.e. $y=\phi(x)$ where $x$ is the drive system and $y$ is the response system. For the unidirectionally coupled systems the evolution of drive system does not depend on the evolution of response system, but for bidirectionally or mutually coupled systems the state variables of each system will depend on the state variables of the other system. So, for such a case the functional relation $y=\phi(x)$ is to be modified to the form~\cite{Cross2002,Moskalenko2012},
\begin{equation}
\psi(x,y) = 0 \label{func-rel-mutual}
\end{equation}

\begin{figure}
\begin{center}
\includegraphics[width=.9\columnwidth]{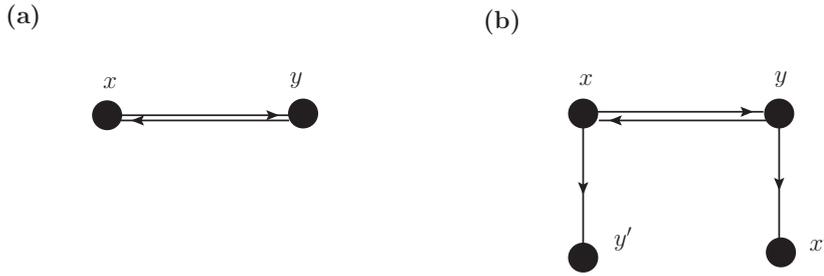}
\end{center}
\caption{(a) The figure shows mutually coupling configuration for two systems, $x$ and $y$.(b) The figure shows the auxiliary system configuration to detect the generalized synchronization between two mutually coupled systems, $x$ and $y$, with $x'$ and $y'$ as the auxilliary systems.}
\label{schematic-mutual-conf}
\end{figure}

In Fig.~\ref{schematic-mutual-conf}(a) the schematic diagram of two mutually coupled systems $x$ and $y$ is shown. Let us consider the dynamics of two mutually coupled systems, $x$ and $y$ given by,
\begin{subequations}
\label{dynamics-mutual}
\begin{eqnarray}
\dot{x} &=& F(x) + \varepsilon_1 H_1(y,x) \label{dyn-mutual-eqn1} \\
\dot{y} &=& G(x) + \varepsilon_2 H_2(x,y) \label{dyn-mutual-eqn2} 
\end{eqnarray}
\end{subequations}
where $x(\in R^d)$ and $y(\in R^r)$ are the state variables of the coupled systems, $\varepsilon_1,\varepsilon_2$ are the coupling parameters and $H_1: R^d \rightarrow R^r$ and $H_2: R^r \rightarrow R^d$ are the coupling functions. For suitable coupling functions and coupling parameters the systems $x$ and $y$ will show GS.

\subsection{ASA to detect GS for mutually coupled systems}
\label{ASA-mutual}

In section \ref{ASA-unidir}, we have discussed the ASA to detect GS for unidirectionally coupled systems. For unidirectionally coupled systems, an auxiliary system of the response system is created by replicating the response system and this auxiliary system is driven with the same drive system. In the stable GS the Euclidean distance between the response system and the auxiliary system goes to zero. 

For mutually coupled systems we need to consider the auxiliary system corresponding to each system and drive these auxiliary systems in the same way as it was done for the original copy. Fig.~\ref{schematic-mutual-conf}(b) shows the schematic diagram for ASA for two mutually coupled systems $x$ and $y$. For Eq.~(\ref{dynamics-mutual}) we consider the following auxiliary systems
\begin{subequations}
\label{aux-sys-mutual}
\begin{eqnarray}
\dot{x}' &=& F(x') + \varepsilon_1 H_1(y,x') \label{aux-dyn-mutual-eqn1}\\
\dot{y}' &=& G(y') + \varepsilon_2 H_2(x,y') \label{aux-dyn-mutual-eqn2}
\end{eqnarray}
\end{subequations}
where $x'(\in R^d)$ and $y'(\in R^r)$ are the state variables of the auxiliary systems. Let $D_x$ and $D_y$ give the time average Euclidean distances between the systems $x$ and $x'$ and the systems $y$ and $y'$ respectively. In the GS, the Euclidean distance between the auxiliary system and its original copy will go to zero.

Let us demonstrate the ASA method with the help of two mutually coupled nonidentical R\"ossler systems
\begin{eqnarray}
\dot{x}_1 &=& -\omega_x x_2 + x_3 + \varepsilon(y_1 - x_1)\nonumber\\
\dot{x}_2 &=& \omega_x x_1 + a_r x_2\label{mutual-two-ros-sys-x} \\
\dot{x}_3 &=& b_r + x_3(x_1 - c_r)\nonumber \\
\nonumber \\
\dot{y}_1 &=& -\omega_y y_2 + y_3 + \varepsilon(x_1 - y_1)\nonumber\\
\dot{y}_2 &=& \omega_y y_1 + a_r y_2\label{mutual-two-ros-sys-y} \\
\dot{y}_3 &=& b_r + y_3(y_1 - c_r)\nonumber
\end{eqnarray}
In Fig.~\ref{err_re_aux_ale_ros_mutual}(a) the time average Euclidean distances $D_x$ and $D_y$ are plotted as a function of the coupling parameter $\varepsilon$. As the coupling parameter $\varepsilon$ is increased the Euclidean distance between systems $y$ and $y'$ goess to zero first, while the $D_x$ is still nonzero. At $\varepsilon=\varepsilon_{GS}\sim 0.087$, $D_x$ goes to zero and at this point 
 the coupled systems undergo generalized synchronization \cite{note-ws}.

\subsection{LE method to analyze GS for mutually coupled systems}
\label{CLE-mutual}

In section \ref{lyap_expo} we have discussed the LE method to detect GS for unidirectionally coupled systems. LE method provides the exact boundary of synchronization both for CS~\cite{Pecora1990,Pecora1998} and for GS~\cite{Parlitz1996}. In the synchronized state the largest transverse Lyapunov exponent (TLE) is negative. We use this criteria the analyze stability of the GS for coupled nonidentical systems.

\begin{figure}
\begin{center}
\includegraphics[width=.9\columnwidth]{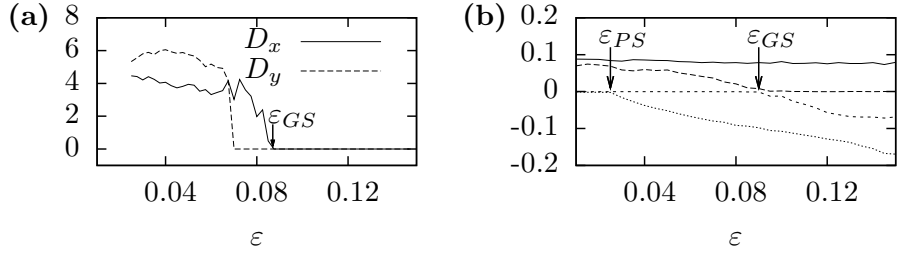}
\end{center}
\caption{(a) The time average Euclidean distance between the auxiliary systems and their respective original systems, $D_x$ and $D_y$, are shown as a function of the coupling parameter $\varepsilon$. AT $\varepsilon=\varepsilon_{GS}\sim 0.087$ both $D_x$ and $D_y$ go to zero and the coupled systems show generalized synchronization. (b) The four largest Lyapunov exponents are plotted as a function of the coupling parameter $\varepsilon$. 
As $\varepsilon$ is increased one zero exponent first become negative at $\varepsilon=\varepsilon_{PS}$, and the coupled systems are phase synchronized.  Further, a positive Lyapunov exponent become negative at $\varepsilon=\varepsilon_{GS}\sim0.087$, and the coupled systems undergo generalized synchronization.
For both plots the R\"ossler parameters are $a_r=0.2,b_r=0.2,c_r=7.0$ and $\omega_x=1.015$ and $\omega_y=0.085$.}
\label{err_re_aux_ale_ros_mutual}
\end{figure}

Let us demonstrate the LE method to analyze GS for mutually coupled systems with the help coupled nonidentical R\"ossler systems, Eqs.~(\ref{mutual-two-ros-sys-x}) and~(\ref{mutual-two-ros-sys-y}). In Fig.~\ref{err_re_aux_ale_ros_mutual}(b) the four largest Lyapunov exponents are shown as a function of the coupling parameter $\varepsilon$. As $\varepsilon$ is increased one zero exponent first become negative at $\varepsilon=\varepsilon_{PS}$, and the coupled systems are phase synchronized. When couping parameter is further increased, a positive Lyapunov exponent become negative at $\varepsilon=\varepsilon_{GS}\sim0.087$, and the coupled systems undergo generalized synchronization.

\section{Generalized synchronization in networks}
\label{gen_sync_comp_net}

Recently, the studies of generalized synchronization in complex networks have received much attention~\cite{Moskalenko2012,Hu2008,Guan2009}. We consider the the following network of $N$ coupled systems
\begin{equation}
\dot{x}^i = F(x^i) + \varepsilon\sum_{j=1}^N a_{ij} H(x^j, x^i); \; i=1,...,N
\label{n-coupled-systems}
\end{equation}
where $x^i(\in R^d)$ is the state variable of system $i$, $F: R^d \rightarrow R^d$ gives the dynamics of an isolated system, $H: R^d \rightarrow R^d$ gives the coupling function, $\varepsilon$ is coupling parameter and $A=[a_{ij}]$ is the coupling matrix. Here, we take $A$ to be the adjacency matrix, i.e. $a_{ij}=1$ if the nodes $i$ and $j$ are coupled and zero otherwise. For suitable coupling function and coupling parameter values the coupled systems will show GS. In the stable GS, the state variables of the coupled systems will be related, thus we can write a generic function giving a functional relation between the state variables of all the coupled system as,
\begin{equation}
\Psi(x^1,x^2,...,x^N) = 0. \label{func-rel-network}
\end{equation}

\subsection{Auxiliary System Approach}
Now, we discuss the auxiliary system approach (ASA) to detect GS between the coupled systems given in Eq.~(\ref{n-coupled-systems}). The auxiliary system are created by replicating each coupled system and they are driven exactly in the same way as their original copy is driven. Let $x_i'$ denote the auxiliary system for system $i$, so we have
\begin{eqnarray}
(\dot{x}^i)' = F((x^i)') + \varepsilon \sum_{j=1}^N a_{ij} H(x^j, (x^i)');\; j=1,..,N
\label{i-auxiliary-systems}
\end{eqnarray}

As an example, we consider a network of $N=8$ randomly coupled R\"ossler systems
\begin{eqnarray}
\dot{x}^i &=& -\omega_i y^i - z^i + \varepsilon \sum_{j=1}^N a_{ij}(x^j - x^i)\nonumber\\
\dot{y}^i &=& \omega_i x^i + a_r y^i \label{8-coupled-ros-sys}\\
\dot{z}^i &=& b_r + z^i(x^i-c_r) \nonumber
\end{eqnarray}
The non-identity between the coupled R\"ossler systems is introduced through the parameter $\omega_i$.

\begin{figure}
\centering
\includegraphics[]{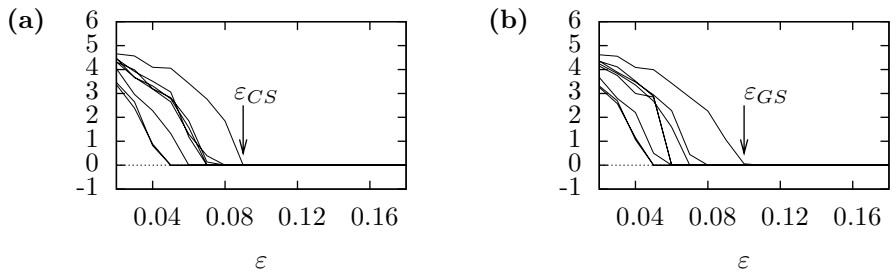}
\caption{(a) The Euclidean distances between the auxiliary systems and their original systems are shown as a function of the coupling parameter $\varepsilon$ for $8$ coupled identical R\"ossler systems. At $\varepsilon=\varepsilon_{CS} = 0.087$ all distances go to zero, indicating a transition to complete synchronization. (b) The Euclidean distances between the auxiliary systems and their original systems are shown as a function of the coupling parameter $\varepsilon$ for $8$ coupled nonidentical R\"ossler systems. At $\varepsilon=\varepsilon_{GS} = 0.10$ all distances go to zero, indicating a transition to generalized synchronization. The R\"ossler parameter $\omega_i$ are chosen randomly from the interval $(.99,1.01)$. Other R\"ossler parameters are $a_r=b_r=0.2,c_r=7$.}
\label{err_res_aux_cs_gs_8sys}
\end{figure}

First, we consider the case when the coupled systems are identical, i.e. $\omega_i = \omega; \; \forall i$. Fig.~\ref{err_res_aux_cs_gs_8sys}(a) shows the Euclidean distances $D_i$ between each auxiliary system and its original, as a function of the coupling parameter $\varepsilon$ for the case when all R\"ossler systems are identical. As the coupling parameter $\varepsilon$ increases some distances go to zero while others remain nonzero. At $\varepsilon=\varepsilon_{CS}$ all distances go to zero, indicating a transition to complete synchronization. A similar behavior is observed for coupled nonidentical systems. Fig.~\ref{err_res_aux_cs_gs_8sys}(b) shows the Euclidean distances $D_i$ between each auxiliary system and its original, as a function of the coupling parameter $\varepsilon$ for the case when the R\"ossler systems are nonidentical. As the coupling parameter $\varepsilon$ increases some distances go to zero and at $\varepsilon=\varepsilon_{GS}$ all distances go to zero, indicating a transition to 
generalized synchronization. 

\subsection{Lyapunov Exponent method}

For networks of coupled identical systems the stability of complete synchronization has been well analyzed. Pecora and Carroll (1998)~\cite{Pecora1998} introduced a master stability function (MSF) which can be calculated from a simple set of master stability equations. Using the master stability function one can calculate the largest transverse Lyapunov exponent for a network. For stable CS,  the largest transverse Lyapunov exponent is negative.

\subsection{\label{MSF} Master Stability Function for coupled nearly-identical systems}
\label{MSF-approach}

In Ref.~\cite{Acharyya2012}, we extend the formalism of MSF to coupled nearly-identical systems. In this section we briefly review the analysis of MSF for coupled nearly-identical systems. We start by considering a network of $N$ coupled dynamical systems as
\begin{equation}
\dot{x}^i = f(x^i,r^i) + \varepsilon\sum_{j=1}^N g_{ij} h(x^j);\; i=1,...,N 
\label{N-systems}
\end{equation}
where $x^i(\in R^m)$ is the $m$-dimensional state vector of system $i$ and $r^i$ is the parameter which makes the systems nonidentical, $f:R^m \rightarrow R^m$ and $h: R^m \rightarrow R^m$ give respectively the dynamical evolution of a single system and the coupling function, $G=[g_{ij}]$ is the coupling matrix and $\varepsilon$ is the coupling constant. The coupling matrix $g_{ij}=1$ when system $i$ couples with system $j$, otherwise $g_{ij}=0$; the diagonal element $g_{ii}= -k^i$, where $k^i$ is the degree of system $i$. So, the coupling matrix satisfies the condition $\sum_{j}g_{ij}=0$ which fulfills the condition for invariance of the synchronization manifold~\cite{Pecora1998}. Let the parameter $r^i = \tilde{r}+\delta r^i$, where $\tilde{r}$ is some typical value of the parameter and $\delta r^i$ is a small mismatch.

When all coupled systems are identical, i.e. $r^i = r;\; \forall i$, the coupled systems exhibit complete synchronization for suitable coupling constant $\varepsilon$~\cite{Pecora1990}. For complete synchronization all state variables of the coupled systems become equal, i.e. $x^i=x;\; \forall i$ and the motion of the coupled systems are confined to a subspace which is the synchronization manifold. The synchronized state is stable when all the transverse Lyapunov exponents are negative. For coupled identical systems the linearized equations can be obtained from Eq.~(\ref{N-systems}) by expanding in Taylor's series about the complete synchronized state, i.e. $x^i = x;\; \forall i$. These linearized equations can be diagonalized into $N$ modes~\cite{Pecora1998} and can written in the form
\begin{equation}
\dot{\phi}^k = [D_x f(x,r) + \varepsilon \gamma_k D_x h(x)] \phi^k \label{pecora-MSF-eq}
\end{equation}
where $D_x$ is the differential operator and $\gamma_k$ is the $k$-th eigenvalue of coupling matrix $G$. Eq.~(\ref{pecora-MSF-eq}) is called as the master stability equation~\cite{Pecora1998}. The MSF is calculated as the largest Lyapunov exponent of Eq.~(\ref{pecora-MSF-eq}) as a function of the parameter $\alpha = \varepsilon\gamma_k$ .

For coupled nonidentical systems, the synchronization will be of the generalized type, where the state variables of the coupled systems are related by a functional relationship~\cite{Abarbanel1995}. For coupled nonidentical systems, it is not possible to have a simple block diagonalized form as in Eq.~(\ref{pecora-MSF-eq}). In Ref.\cite{Acharyya2012} we have shown how one can achieve an approximate block diagonalized form similar to Eq.~(\ref{pecora-MSF-eq}) for nearly-identical systems. Using this form we can obtain the master stability function and thus determine the stability of the generalized synchronization for nearly-identical systems.

For coupled identical systems the variational equations are obtained by expanding Eq.~(\ref{N-systems}) around the synchronous solution $x^i = x;\; \forall i$, where the $x$ can be obtained by integrating an isolated system. Similar expansion is not possible when we consider the case of coupled nearly-identical systems, as now the synchronous solution is not the solution of an isolated system, but it is given by some functional relation between the state variables of coupled systems. One way of doing this expansion is to consider the average trajectory $\bar{x}=1/N\sum_i x^i$, and expand Eq.~(\ref{N-systems}) around this average trajectory~\cite{Sun2009}. This needs the integration of all the systems and the computation increases for large networks. Other way of doing this is to expand Eq.~(\ref{N-systems}) around the solution of an isolated system with some typical parameter $\tilde{r}$. In the generalized synchronization, the state variables of the coupled systems are related with some functional relation. Hence, we can assume that the attractors of the coupled systems are not much different from each other so that the average rate of expansion and contraction for attractors of the coupled systems are also not very different. One choice of typical parameter value is the average value, i.e. $\tilde{r} = \bar{r}= 1/N \sum_i r^i$, and it gives a good approximation to the Lyapunov exponents (see Figure 1 of Ref.~\cite{Acharyya2012}). Thus, we expand Eq.~(\ref{N-systems}) in Taylor's series about the solution $\tilde{x}$ of an isolated systems with typical parameter value $\tilde{r}$. We retain terms up-to second order and we get
\begin{eqnarray}
\dot{z}^i &=& D_x f(\tilde{x},\tilde{r}) z^i + D_r f(\tilde{x},\tilde{r}) \delta r^i + \frac{1}{2} D_r^2 f(\tilde{x},\tilde{r}) (\delta r^i)^2\nonumber \\ 
&&  + D_r D_x f(\tilde{x},\tilde{r}) z^i \delta r^i  + \varepsilon \sum_{j=1}^N g_{ij} D_x h(\tilde{x}) z^j 
\label{first-taylor-expand}
\end{eqnarray}
where $z^i = x^i - \tilde{x}$ is the deviation of $i$-th system from $\tilde{x}$. In Eq.~(\ref{first-taylor-expand}) we have dropped the term containing $(z^i)^2$ as we are interested in the solution $z^i \rightarrow 0$. Eq.~(\ref{first-taylor-expand}) contains both inhomogeneous and homogeneous terms. The exponential dependence of solutions of a linear differential equation are given by the homogeneous terms~\cite{Acharyya2012,Sorrentino2011}. So to calculate Lyapunov exponents from Eq.~(\ref{first-taylor-expand}) we can drop the inhomogeneous terms to obtain
\begin{equation}
\dot{z}^i = D_x f(\tilde{x},\tilde{r}) z^i + D_r D_x f(\tilde{x},\tilde{r}) z^i \delta r^i + \varepsilon\sum_{j=1}^N g_{ij} D_x h(\tilde{x})
\label{first-homogeneous-equation}
\end{equation}
Eq.~(\ref{first-homogeneous-equation}) can be put in matrix form as
\begin{equation}
\dot{Z} = D_x f(\tilde{x},\tilde{r}) \ Z + D_r D_x f(\tilde{x},\tilde{r}) \ Z \ R + D_x h(\tilde{x}) \ Z \ G^T
\label{first-matrix-form}
\end{equation}
where $G^T$ is the transpose of the coupling matrix $G$ and $Z=(z^1,..,z^N)$ and $R={\rm diag}(\delta r^1,...,\delta r^N)$. Now, we want to decouple Eq.~(\ref{first-matrix-form}) along the eigenvalues of the coupling matrix $G^T$. Let $\gamma_j,\;j=1,..N$ be the eigenvalues of the coupling matrix $G^T$ and the corresponding left and right eigenvectors are $e_j^L$ and $e_j^R$ respectively. We note that there exists an eigenvalue $\gamma_1=0$ of the coupling matrix $G^T$ and it defines the synchronization manifold and the rest of the eigenvalues define the transverse manifold. We multiply Eq.~(\ref{first-matrix-form}) by $e_j^R$ from right and use the $m$-dimensional vector $\phi_j=Z e_j^R$. Thus,
\begin{equation}
\dot{\phi}_j = [D_x f + \varepsilon \gamma_j D_x h] \phi_j + D_r D_x f \ Z \ R e_j^R.
\label{second-matrix-form}
\end{equation}
In Eq.~(\ref{second-matrix-form}) $e_j^R$ are not eigenvalues of $R$. Hence, we use first order perturbation theory and obtain the first order correction due to the parameter mismatch as $\nu_j = e_j^L R e_j^R$. Thus, we can approximate Eq.~(\ref{second-matrix-form}) as
\begin{equation}
\dot{\phi}_j = [D_x f + \varepsilon \gamma_j D_x h] \phi_j + \nu_j D_r D_x \phi_j.
\label{first-master-stability-equation}
\end{equation}
The generic variational equation or the master stability equation can be written by considering two complex parameters $\alpha = \varepsilon\gamma_j$ and $\Delta=\nu_j$ as
\begin{equation}
\dot{\phi} = [D_x f + \alpha D_x h + \Delta D_r D_x f]
\label{master-stability-equation}
\end{equation}
The master stability function (MSF) $\lambda_{max}$ is defined as the largest Lyapunov exponent as a function of the parameters $\alpha$ and $\Delta$. The accuracy of the MSF for coupled nearly-identical systems is discussed in Ref~\cite{Acharyya2012}. 

\begin{figure}
\begin{center}
\includegraphics[]{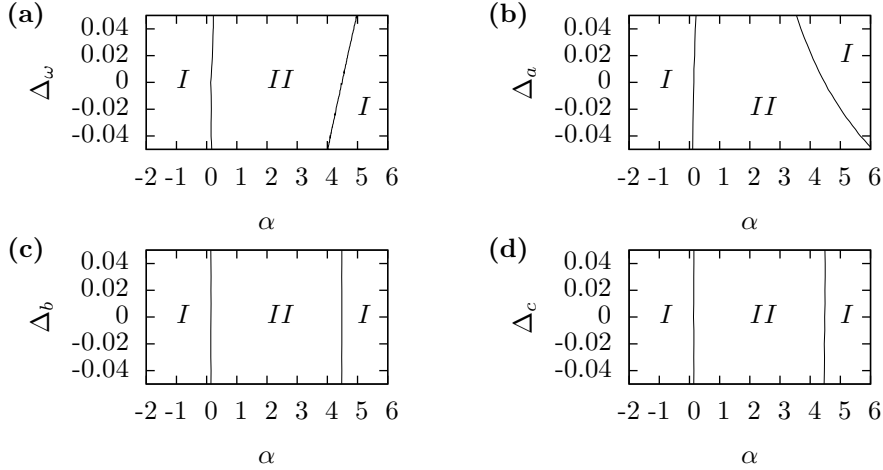}
\end{center}
\caption{The zero contours of the master stability function of $x$ coupled R\"ossler systems are plotted in the parameter plane $\alpha-\Delta$ for four cases, (a) mismatch in parameter $\omega$, (b) mismatch in parameter $a_r$, (c) mismatch in parameter $b_r$, and (d) mismatch in parameter $c_r$. MSF is negative in the region bounded by the two zero contour curves. In all the figures region MSF is positive in region $I$ and this region gives the unstable region and MSF is negative in region $II$, thus these regions give the stable region. The R\"ossler parameters are $\omega=1,a_r=b_r=0.2,c_r=7.0$.}
\label{msf_ros}
\end{figure}

Now we demonstrate the MSF by considering coupled nearly-identical R\"ossler systems. The dynamics of a single R\"ossler system is given by
\begin{eqnarray}
\dot{x} &=& -\omega y - z \nonumber \\
\dot{y} &=& \omega x + a_r y \label{rossler-system}\\
\dot{z} &=& b_r + z(x - c_r) \nonumber
\end{eqnarray}
We consider that the systems are coupled in the $x$ component.  Let us consider the simpler case of R\"ossler systems with mismatch in one parameter only. The zero contour curves of the master stability function for this case are given in Fig.~\ref{msf_ros} in the $\alpha-\Delta$ plane with (a),(b),(c), and (d) giving the zero contour curves when the mismatch is present in R\"ossler parameters $\omega, \, a_r, \, b_r,$ and $c_r$ respectively. The MSF is negative in the region II border by the two zero contour curves and is positive in region I outside. We will refer to region I as the unstable region and region II as the stable region.  

From Fig.~\ref{msf_ros}(a) we can see that the stable region increases with increase in the parameter $\Delta_{\omega}$ and  from Fig.~\ref{msf_ros}(b) that the stable region decreases with decrease in parameter $\Delta_a$. Figs.~\ref{msf_ros}(c) and (d) show that the stability of the GS is almost unaffected for mismatch in parameters $b_r$ and $c_r$.

Let us consider a specific network of $N$ coupled R\"ossler systems with coupling matrix $G$ and mismatch in one parameter. Determine the eigenvalues $\gamma_i, i=1,\ldots,N$ of $G$. There is one eigenvalue $\gamma_1=0$ which corresponds to the synchronization manifold. For the other eigenvalues, determine the $N-1$ pairs of parameters $\alpha_i$ and $\Delta_i$. If for all these $N-1$ pairs of parameters the MSF lies in the stable region then the coupled systems are in stable GS.

\section{Synchronized optimized networks}
\label{sync-opt-net}

The MSF for coupled chaotic systems can be used to construct synchronized optimized networks.  By synchronized optimized network we mean that the synchronization is stable for the widest possible interval in the coupling parameter. Let us consider a network of coupled nearly-identical R\"ossler systems with mismatch in parameter $\omega$. From Fig.~\ref{msf_ros}(a) we can see that the stability region increases with increase in the mismatch parameter $\Delta_{\omega}$. Now to construct synchronized optimized networks from this given network we rewire its link and we accept the new network if the stable interval increases, otherwise we accept the network with probability $p=\exp{(l_{new}-l_{old})/T}$, where $l_{old}$ and $l_{new}$ are the stable intervals of the network before and after rewiring and $T$ is a temperature like quantity. The temperature $T$ is reduced after certain number of iterations so that simulated annealing occurs. We stop the optimization method when there is no change in the network for 
five successive temperature steps. At this point we assume that a good approximation of the optimal network has been achieved.

When we construct synchronized optimized networks from coupled nonidentical systems then there are additional questions such as which nodes become hubs of the network and which links are more preferred in the optimal network. To answer these questions we introduce the following correlations coefficients. We define the average correlation coefficient between the degree of a node and its parameter as
\begin{equation}
\rho_{rk} = \frac{1}{N} \sum_i \frac{\langle(r^i - \langle r^i\rangle)(k^i - \langle k^i\rangle)\rangle}{\sqrt{\langle(r^i - \langle r^i\rangle)^2\rangle\langle(k^i - \langle k^i\rangle)^2\rangle}}
\label{param-deg-corr}
\end{equation}
where $k^i=-g_{ii}$ is the degree of node $i$ and $r^i$ is its parameter. For a random network $\rho_{rk}=0$. Next, to find which links are more preferred in the optimal networks we define the average correlation between the parameter difference between a pair of nodes and their connection as
\begin{equation}
\rho_{rA} = \frac{1}{N} \sum_i \frac{\langle(\delta r^{ij} - \langle\delta r^{ij}\rangle)(a_{ij} - \langle a_{ij}\rangle)\rangle}{\sqrt{\langle(\delta r^{ij} - \langle\delta r^{ij}\rangle)^2\rangle\langle(a_{ij} - \langle a_{ij}\rangle)^2\rangle}}
\label{paramdiff-adj-corr}
\end{equation}
where $\delta r^{ij} = |r^i - r^j|$ is the parameter difference between node $i$ and node $j$ and $A=[a_{ij}]$ is the adjacency matrix and $a_{ii}=0$. For a random network $\rho_{rA}=0$.

\begin{figure}
\centering
\includegraphics[]{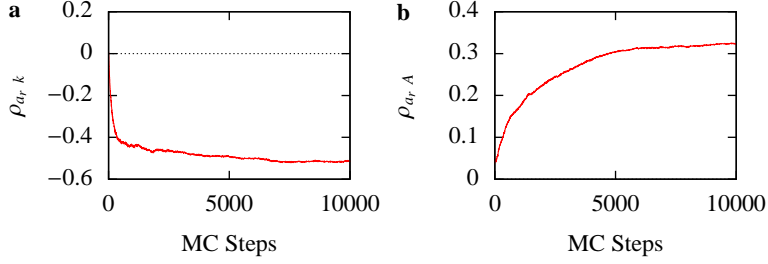}
\caption{(a) The correlation coefficient $\rho_{a_r k}$ is shown as a function of the Monte Carlo steps. This correlation coefficient starts from zero and then saturates to a negative value. (b) The correlation coefficient $\rho_{a_r A}$ is plotted as a function of the Monte Carlo steps. This correlation coefficient increases as a function of Monte Carlo steps and saturates to a positive value.}
\label{corr-ark-arA}
\end{figure}

To see the uses of these correlation coefficients let us consider a network of $N=32$ nearly-coupled R\"ossler systems which have parameter mismatch in the R\"ossler parameter $a_r$. From Fig.~\ref{msf_ros}(b) we can see that for this configuration the stable region increase with decrease in the first order correction term $\Delta_a$. In Fig.~\ref{corr-ark-arA}(a) we plot $\rho_{a_r k}$ as a function of Monte Carlo steps. We can see $\rho_{a_r k}$ starts near zero value and then decreases and saturates to a negative value. This implies that the nodes with smaller value of parameter are likely to have higher degree in the optimal network and thus become the hubs of the optimal network. Fig.~\ref{corr-ark-arA}(b) shows the correlation coefficient $\rho_{a_r A}$ as a function of Monte Carlo steps. $\rho_{a_r A}$ increases from zero and saturates to a positive value. It implies that the pair of nodes which have higher parameter difference are preferred for creating links of the optimal network.

\section{Conclusion}

To conclude we have briefly discussed some recent developments in the study of GS. We discuss different methods of detecting GS such as mutual false nearest neighbors, auxiliary system approach and the Lyapunov exponents method. We have analyzed the stability of the GS for coupled nearly-identical systems with the help of MSF. Using these MSF we later discuss the problem of constructing synchronized optimized networks from a given network with fixed number of links and nodes. We rewire the links to achieve the optimal network. For the synchronized optimized networks we have found that the nodes with parameter value at one extreme are chosen as hubs and the pair of nodes with relatively large parameter difference are chosen to create links.

%
%


\begin{thebibliography}{}
\bibitem{note-ws} Between the region when $D_y$ is zero, but $D_x$ is nonzero, we get weak generalized synchronization. We do not discuss the details of this phenomena here which can be found in
K. Pyragas, Phys. Rev. E \textbf{54}, (1996) R4508.
\end{thebibliography}
\end{document}